\documentclass[12pt,preprint]{aastex}

\shorttitle{$Z$ and $E_{\gamma,iso}$ in LGRBs}
\shortauthors{Levesque et al.}
\slugcomment{DRAFT \today}

\begin{document}
\title{No Correlation Between Host Galaxy Metallicity and Gamma-Ray Energy Release for Long-Duration Gamma-Ray Bursts}
\author{Emily M. Levesque$^{1,2,3}$, Alicia M. Soderberg\footnotemark[2], Lisa J. Kewley\footnotemark[1], Edo Berger\footnotemark[2]}
\email{emsque@ifa.hawaii.edu, asoderbe@cfa.harvard.edu, eberger@cfa.harvard.edu, kewley@ifa.hawaii.edu}

\footnotetext[1]{Institute for Astronomy, University of Hawaii, 2680 Woodlawn Dr., Honolulu, HI 96822}
\footnotetext[2]{Smithsonian Astrophysical Observatory, 60 Garden St., MS-20, Cambridge, MA 02138}
\footnotetext[3]{Current address: CASA, Department of Astrophysical and Planetary Sciences, University of Colorado, 389 UCB, Boulder, CO 80309}

\begin{abstract}
We compare the redshifts, host galaxy metallicities, and isotropic ($E_{\gamma,iso}$) and beaming-corrected ($E_{\gamma}$) gamma-ray energy release of 16 long-duration gamma-ray bursts (LGRBs) at $z < 1$. From this comparison, we find no statistically significant correlation between host metallicity and redshift, $E_{\gamma,iso}$, or $E_{\gamma}$. These results are at odds with previous theoretical and observational predictions of an inverse correlation between gamma-ray energy release and host metallicity, as well as the standard predictions of metallicity-driven wind effects in stellar evolutionary models. We consider the implications that these results have for LGRB progenitor scenarios, and discuss our current understanding of the role that metallicity plays in the production of LGRBs.
\end{abstract}

\section{Introduction}
Gamma-ray bursts (GRBs) are among the most energetic phenomena observed in the universe. The long-duration (T$_{90} >$ 2 s) subclass of these events, LGRBs, are typically associated with the core-collapse of massive stars (Woosley 1993). However, a detailed picture of the progenitors that produce LGRBs is still unclear. Potential mass stripping mechanisms and engine models for LGRBs include winds (e.g. Yoon et al.\ 2006, Woosley \& Heger 2006), binary mergers (e.g. Podsiadlowski et al.\ 2004, 2010; Fryer \& Heger 2005) and magnetars (e.g. Wheeler et al.\ 2000, Burrows et al.\ 2007, Uzdensky \& MacFadyen 2007, Bucciantini et al.\ 2009, Nagataki 2009).

The most commonly-cited progenitor model for LGRBs was originally proposed by Woosley (1993), and is known as the {\it collapsar} model. Under the classical assumptions of this model, the progenitor is a single rapidly-rotating massive star, which maintains a high enough angular momentum over its lifetime to generate a LGRB from core-collapse to an accreting black hole. The collapsar model has been considered somewhat paradoxical since the progenitors, assumed to be Wolf-Rayet stars, are expected to shed their outer H and He shells during periods of high wind-driven mass loss. This in turn would lead to a loss of the angular momentum that is critical for GRB production in the collapsar model. However, multiple models have shown that massive stars evolving in low-metallicity environments retain sufficient angular momentum to produce a GRB, due to the effects of metallicity on line-driven mass loss rates (e.g. Hirschi et al.\ 2005, Yoon et al.\ 2006, Woosley \& Heger 2006).

In recent years, observations of LGRB host galaxies have generally supported the claim that the progenitors may favor low-metallicity environments. The host galaxies are generally fainter, more irregular, and have lower oxygen abundances than the general galaxy population out to $z \sim 1$ (e.g. Stanek et al.\ 2006; Fruchter et al.\ 2006; Kewley et al.\ 2007; Wainwright et al.\ 2007; Modjaz et al.\ 2008; Kocevski et al.\ 2009; Levesque et al.\ 2010a, 2010b). A number of studies have even proposed the idea of a ``cut-off" maximum host metallicity for LGRB production (Stanek et al.\ 2006, Wolf \& Podsiadlowski 2007, Modjaz et al.\ 2008, Kocevski et al.\ 2009), although the discovery of several high-metallicity LGRB host environments suggest that a low-metallicity cut-off is unlikely (Levesque et al.\ 2010b, 2010c; Graham et al.\ 2009). The high-metallicity host galaxy of the relativistic SN 2009bb presents additional evidence that the central-engine-driven relativistic explosions thought to power LGRBs are not restricted to low-metallicity progenitors (Levesque et al.\ 2010d, Soderberg et al.\ 2010a). Due to the young ages assumed for the massive progenitors of LGRBs ($\le$ 10 Myr; Woosley et al.\ 2002), these measured environmental metallicities can be considered representative of the natal abundance properties of the progenitor stars (see S\`{i}mon-D\`{i}az et al.\ 2006, Hunter et al.\ 2007).

Even with these careful studies of LGRBs and their host environments, the role that metallicity plays in progenitor evolution and GRB production remains unclear. If metallicity does indeed have a direct impact on progenitor properties (such as angular momentum) that are key to producing these high-energy core-collapse events, we would therefore expect metallicity to have some correlation with the explosive properties of LGRBs. A low metallicity environment produces stars with higher helium core masses and faster rotation rates, leading to LGRBs that are expected to be more energetic (e.g. MacFadyen \& Woosley 1999); from these assumptions low metallicities should produce LGRBs with a higher energy release in the gamma-ray regime ($E_{\gamma}$).

Several previous studies have investigated this possibility. Ramirez-Ruiz et al.\ (2002) found a tentative positive correlation between the isotropic energy release ($E_{\gamma,iso}$) and the offset of a GRB from the center of its host galaxy ($r_0$). This offset correlation was proposed to be a potential artifact of a correlation between $E_{\gamma,iso}$ and low metallicity -- chemical abundance gradients have shown that stars at higher $r_0$, in the outskirts of their hosts, have lower metallicities on average (e.g. Zaritsky et al.\ 1994, van Zee 1998, Henry \& Worthey 1999). Since $E_{\gamma,iso}$ is calculated assuming a quasi-spherical GRB explosion geometry, rather than accounting for the expected effects of a potential conical geometry with a narrow opening angle for the GRB jet ($\theta_j$; see Frail et al.\ 2001), this result suggested that low metallicity was associated with either higher $E_{\gamma}$ or narrower GRB jets. 

A comparison between host metallicity and $E_{\gamma,iso}$ for 5 nearby GRBs (z $<$ 0.3) was performed by Stanek et al.\ (2006).  These authors found a correlation between the two quantities. All but one of the bursts in this sample were ``sub-luminous" LGRBs, a potentially unique class of GRB with $E_{\gamma,iso}$ values that are much lower than the general population (e.g. Soderberg et al.\ 2004a, Soderberg 2006). Stanek et al.\ (2006) argued that this correlation supported the idea of a ``threshold" metallicity for producing ``cosmological" GRBs with more typical luminosities, given that the burst with the highest $E_{\gamma,iso}$ in the sample was produced in the host galaxy with the lowest metallicity. However, they also cautioned that this conclusion was speculative due to the small size and sub-luminous nature of their sample.

Wolf \& Podsiadlowski (2007) also investigated the possibility of a trend relating $E_{\gamma,iso}$ and metallicity, performing the same comparison described in Stanek et al.\ (2006) and including a sample of 13 ``cosmological" ($z > 0.2$) LGRBs. They concluded that, while the Stanek et al.\ (2006) relation holds true for the nearby sub-luminous LGRBs, no relation is apparent in the larger sample. However, the metallicies derived for these ``cosmological" LGRBs were extrapolated from the general luminosity-metallicity relation for star-forming galaxies, a relation that LGRB host galaxies are now known to not follow (Kewley et al.\ 2007, Modjaz et al.\ 2008, Levesque et al.\ 2010a).

To perform a robust test for a correlation between metallicity and the gamma-ray energy release, a large and uniform sample of LGRBs with known host metallicities, $E_{\gamma,iso}$, and $\theta_j$ is required. Here we present the results of such a comparison, using LGRB host metallicities from Levesque et al.\ (2010a, 2010b) and $E_{\gamma,iso}$ and $\theta_j$ measurements for these LGRBs drawn from the literature.

\section{Comparing Host Metallicity and Gamma-Ray Energy Release}
We have compiled redshifts, host metallicities, $E_{\gamma,iso}$, and $\theta_j$ values for a sample of 16 $z < 1$ LGRBs. The redshifts were determined from GCN circulars and other current literature (see Table 1). Metallicities for all but one of host galaxies were taken from Levesque et al.\ (2010a,b); for GRB 100316D we use the metallicity from Chornock et al.\ (2010). All metallicities in our sample were determined using the R$_{23}$ metallicity diagnostic presented in Kewley \& Dopita (2002) and refined in Kobulnicky \& Kewley (2004) and Kewley \& Ellison (2008). For the host galaxies of GRBs 980703 and 020405, we include both upper and lower metallicities since the degeneracy in the R$_{23}$ diagnostic could not be resolved (see Levesque et al.\ 2010a,b). Since the majority of the hosts in our sample are unresolved, these global host metallicities are adopted as a proxy for progenitor metallicity in out analyses; it is expected that, in general, a low-metallicity progenitor would be associated with a low-metallicity host. It is also worth noting that most LGRB host galaxies are low-luminosity compact dwarfs, which are known to have small metallicity gradients (however, see also the analysis of the unusual and well-resolved GRB 980425 host galaxy presented in Christensen et al.\ 2008). Our metallicities also differ in many cases from those given for the nearby LGRB hosts presented in Stanek et al.\ (2006); however, these metallicities were determined from a variety of different spectra (Sollerman et al.\ 2005, Hammer et al.\ 2006, Gorosabel et al.\ 2005, Modjaz et al.\ 2006) that included differing resolutions, S/N, contributions from the underlying supernovae, and large uncertainties in flux calibrations. We consider the results of our uniform host galaxy survey to be more robust.

For all but two of the LGRBs in our sample, values for $E_{\gamma,iso}$ were taken from Amati (2006) and Amati et al.\ (2008). In the case of GRB 050826, we adopt the value for $E_{\gamma,iso}$ from Butler et al.\ (2007), and for GRB 100316D we adopt $E_{\gamma,iso}$ from Starling et al.\ (2010). Amati (2006) and Amati et al.\ (2008) determine $E_{\gamma,iso}$ ranging over (10 - 10,000)/(1+$z$) keV, extrapolating from fits to data drawn from various instruments. While Amati (2006) note that this approach could produce systematic errors, the data are internally self-consistent. Butler et al.\ (2007) determined $E_{\gamma,iso}$ in the 1-10,000 keV band for GRB 050826. Amati (2006) estimates that the difference inherent in integrating from 1-10,000 keV rather than 10-10,000 keV integrations is typically on the order of 3-5\%, and no larger than 10\%. Starling et al.\ (2010) find a lower limit for $E_{\gamma,iso}$ in the 1-160 keV range for GRB 100316D ({\it Swift} satellite observations were temporarily halted before $\gamma$-ray emission from the burst had fully ceased). They note that there is very little flux observed for this nearby ($z = 0.059$) sub-luminous GRB above 50 keV; as a result, additional contribution to the isotropic energy from the higher-energy regime is expected to be negligible.

To convert from $E_{\gamma,iso}$ into $E_{\gamma}$, we adopted values for $\theta_j$ taken from the literature. The jet opening angle $\theta_j$ is typically calculated based on the time of an observed ``break" in the afterglow lightcurve and using the formula of Sari et al.\ (1999); in cases where the jet break time is unclear, a lower or upper limit can be placed on $\theta_j$. For GRB 030528, we adopt the Rau et al.\ (2004) approximation of a jet break at $\sim$3 days, and use this to calculate a $\theta_j = 0.14$ radians. For bursts that have shown evidence of a quasi-spherical, rather than conical, explosion geometry (GRBs 980425, 020903, 031203, 060218, and 100316D), $E_{\gamma,iso}$ and $E_{\gamma}$ are roughly equivalent.

We plot host metallicity against redshift, $E_{\gamma,iso}$, and $E_{\gamma}$ in Figure 1. We have included the redshift comparison to illustrate any potential correlation that may appear as an artifact of metallicity evolution with redshift. For host galaxies with both upper and lower branch metallicities from the Kobulnicky \& Kewley (2004) R$_{23}$ diagnostic, we plot two data points connected by a dotted line to illustrate their correspondence to a single host. We find that there is no statistically significant correlation between metallicity and redshift (Pearson's $r = 0.10$, $p = 0.71$ assuming lower-branch metallicities, and Pearson's $r = 0.28$, $p = 0.29$ assuming upper-branch metallicities).

Similarly, when comparing metallicity and $E_{\gamma,iso}$, we again find no statistically significant correlation (Pearson's $r = 0.08, p = 0.78$ assuming lower-branch metallicities, and Pearson's $r = 0.10, p = 0.72$ assuming upper-branch metallicities). This result is at odds with the inverse correlation between $E_{\gamma,iso}$ and metallicity proposed by Stanek et al.\ (2006). Finally, we find no statistically significant correlation between metallicity and $E_{\gamma}$ (Pearson's $r = 0.18, p = 0.60$ assuming lower-branch metallicities, and Pearson's $r = 0.37, p = 0.26$ assuming upper-branch metallicities). Therefore, even taking beaming effects into account we find no evidence for a relation between host metallicity and the true gamma-ray energy release in LGRBs

\section{Discussion}
The complexities of the role that metallicity plays in the production of LGRBs and the evolution of their progenitors can be summarized by several key findings:

\begin{enumerate}
\item{From a comparison of 16 $z < 1$ LGRBs in this work, we find no evidence for a correlation between host metallicity and gamma-ray energy release, considering both $E_{\gamma,iso}$ and $E_{\gamma}$. This conclusion does not agree with the predictions and claimed relations of past work (e.g. MacFadyen \& Woosley 1999, Ramirez-Ruiz et al.\ 2002, Stanek et al.\ 2006). This is also at odds with the standard predictions of metallicity-driven wind effects in LGRB progenitor evolutionary models.}
\item{There does not appear to be a clear cut-off host metallicity for LGRBs, as has been previously proposed (e.g. Wolf \& Podsiadlowski 2007, Modjaz et al.\ 2008, Kocevski et al.\ 2009); we have observed LGRBs in high-metallicity environments (Graham et al.\ 2009; Levesque et al.\ 2010b, 2010c). There does not yet appear to be a clear maximum metallicity above which LGRB progenitors cannot be formed.}
\item{A number of studies demonstrate that LGRBs {\it do} preferentially occur in galaxies with lower metallicities than the general population. LGRB hosts observed out to $z \sim 1$ fall below the luminosity-metallicity and mass-metallicity relations for the general star-forming galaxy population (e.g. Stanek et al.\ 2006, Kewley et al.\ 2007, Modjaz et al.\ 2008, Levesque et al.\ 2010a,b). However, the physical phenomena driving this trend remain unclear.}
\end{enumerate}

The key assumptions that are refuted by items 1 and 2 -- a correlation between lower host metallicities and higher gamma-ray energy releases, and a proposed upper metallicity cut-off for LGRB host galaxies -- are based on the traditional collapsar model and current assumptions regarding the effects of metallicity on massive star evolution. Specifically, under the assumption of the collapsar model, lower-metallicity host environments are expected to produce progenitors with higher angular momentum, which should in turn produce LGRBs with higher gamma-ray energy releases (e.g. MacFadyen \& Woosley 2001, Hirschi et al.\ 2005, Yoon et al.\ 2006, Woosley \& Heger 2006). However, current evolutionary models for massive stars do not properly address the difficulties of modeling mass loss mechanisms, which may be anisotropic (Meynet \& Maeder 2007) and could also include complex effects such as wind clumping (Crowther et al.\ 2002) and rotation-driven mass loss effects (Meynet \& Maeder 2000). According to Dessart et al.\ (2008), current magnetohydrodynamic simulations of low-metallicity massive stars actually produce core angular momenta that are too high to generate GRB-producing collapsars. Adopting complete and rigorous treatments of mass loss components and magnetic processes in massive stellar evolutionary models could potentially yield evolutionary pathways for collapsars that are not strongly dependent on a strict cut-off for the progenitor's natal metallicity. It is also possible that some of the variation seen in our comparison could be attributed to variations in the initial masses of the LGRB progenitor stars. We are not able to extrapolate any information about stellar progenitor masses from the data. However, it is clear that metallicity is not the sole determinant of $E_{\gamma,iso}$ or $E_{\gamma}$ in LGRB production, and that additional parameters must be considered in future studies.

Alternatively, it is also possible that these recent results regarding LGRBs and their host metallicities may be in better agreement with other alternative progenitor pathways, such as magnetar or binary scenarios. These models do not necessarily require a low-metallicity environment for the evolution and development of the critical mechanism that produces a LGRB (see, for example, Fryer \& Heger 2005; Podsiadlowski et al.\ 2010). However, it is not yet clear whether any of these possibilities can adequately address the phenemonon observed in item 3. We cannot yet provide a physical explanation for why LGRB hosts would only have lower metallicities {\it relative to their mass}, rather than metallicities that are uniformly low or fall below a particular threshold, although this may relate to the young progenitor ages and star formation histories of the hosts (see Berger et al.\ 2007). For the moment, it is clear that, while a lower-than-average host metallicity is a key component of LGRBs, the role of metallicity in progenitor evolution and LGRB production currently remains a mystery. A number of high-redshift host galaxies currently have absorption-line metallicities available in the literature (e.g. Prochaska et al.\ 2007, Savaglio et al.\ 2009, Rau et al.\ 2010); while the lack of satisfactory calibrations between emission-line and absorption-line metallicities precludes adding these high-redshift galaxies to the current sample, future extension of this work to include a greater redshift range of LGRB hosts would be valuable. Subsequent work in this area would also benefit greatly from an improved understanding, both observational and theoretical, of the various mechanisms that drive mass loss and impact late-type evolution in massive stars.

We thank the anonymous referee for their valuable feedback on this manuscript. We gratefully acknowledge useful discussions with David Pitman and Mike Shull. This paper made use of data from the Gamma-Ray Burst Coordinates Network (GCN) circulars and the {\it Swift} online data archive. E.M.L.'s participation was made possible in part by a Ford Foundation Predoctoral Fellowship and an Einstein Fellowship. A.M.S. is supported by a Hubble Fellowship. L.J.K. and E.M.L. acknowledge support by NSF EARLY CAREER AWARD AST07-48559. E. B. acknowledges support by NASA/Swift AOJ grant 5080010.

\begin{figure}
\epsscale{1}
\plotone{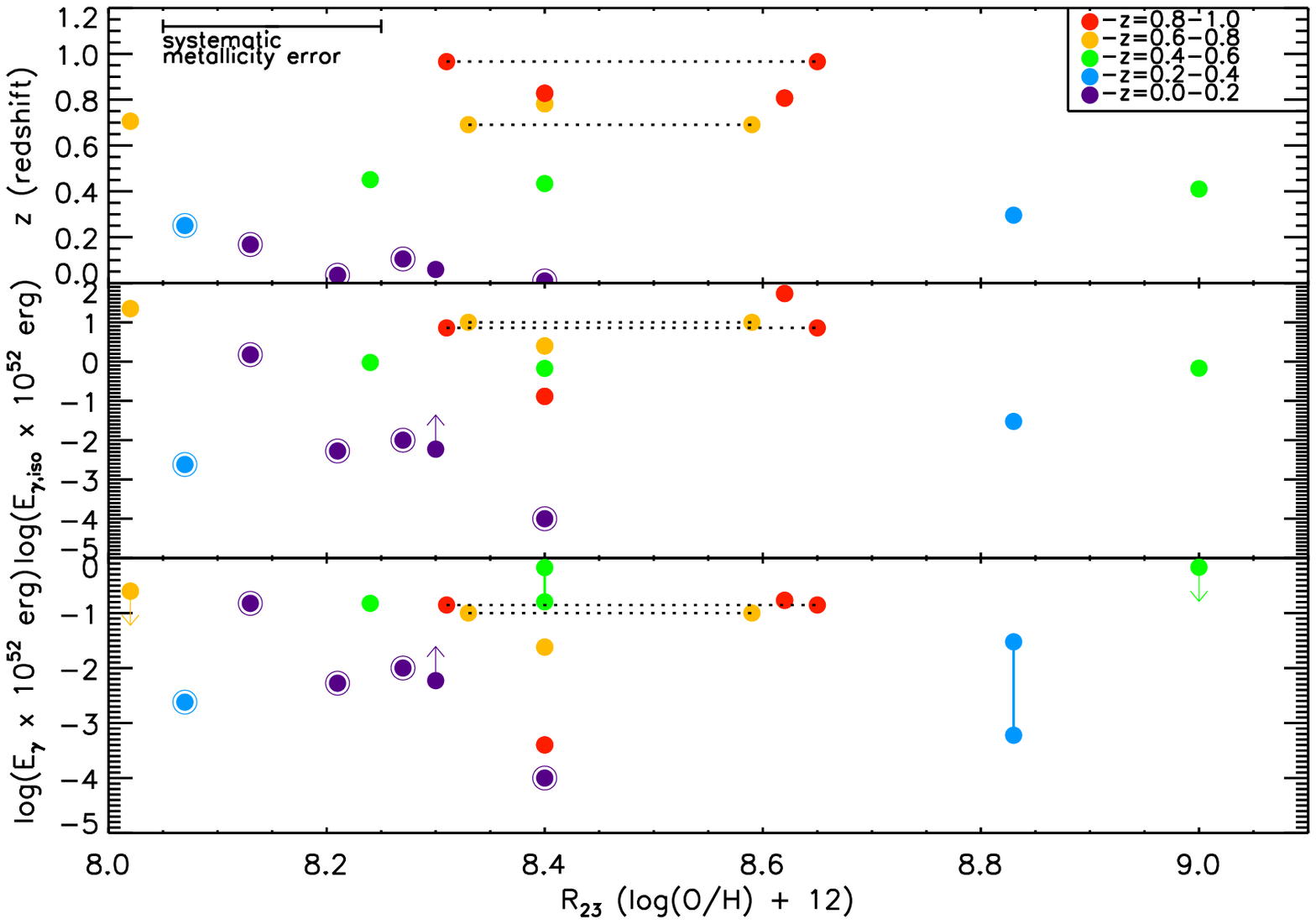}
\caption{The metallicity vs. redshift (top), $E_{\gamma,iso}$ (center), and $E_{\gamma}$ relations for our sample of LGRB host galaxies. The hosts have been separated into redshift bins by color in order to better illustrate redshift effects in these comparisons. Host galaxies with both lower- and upper-branch metallicities from the Kobulnicky \& Kewley (2004) R$_{23}$ diagnostic are indicated by lower and upper data points connected by dotted lines. Upper and lower limits are indicated by arrows. Host galaxies with both upper {\it and} lower limits on their $E_{\gamma}$ values are indicated by data points connected by solid lines. The five nearby LGRB host galaxies included in the original Stanek et al.\ (2006) relation are marked with outer circles.}
\end{figure}

\clearpage
\begin{deluxetable}{l c c c c c c}
\tabletypesize{\scriptsize}
\tablewidth{0pc}
\tablenum{1}
\tablecolumns{7}
\tablecaption{\label{tab:gals} Properties of Nearby ($z < 1$) LGRBs}
\tablehead{
\colhead{GRB}
&\colhead{$z$}
&\colhead{log(O/H) + 12\tablenotemark{a}}
&\colhead{E$_{\gamma, iso}$ (10$^{52}$ erg)}
&\colhead{$\theta_j$ (radians)}
&\colhead{E$_{\gamma}$ (10$^{52}$ erg)}
&\colhead{References\tablenotemark{b}}
}
\startdata
980425 & 0.009   &$\sim$8.4$^{+0.2}_{-0.1}$ &0.00010 $\pm$ 0.00002 &quasi-spherical & 0.00010 $\pm$ 0.00002 &1,2,3,4,5\\ 
060218 &0.034  &8.21 &0.0053 $\pm$ 0.0003 &quasi-spherical &0.0053 $\pm$ 0.0003 &6,2,7,8 \\
100316D &0.059 &8.3 &$\ge$ 0.0059 $\pm$ 0.0005 &quasi-spherical &$\ge$ 0.0059 $\pm$ 0.0005 &9,10,11,12\\
031203 &0.105   &8.27 &0.010 $\pm$ 0.004 &quasi-spherical &0.010 $\pm$ 0.004 &13,2,3,14 \\ 
030329 &0.168  &8.13 &1.5 $\pm$ 0.3 &0.45 &0.15 $\pm$ 0.03  &15,2,7,16,17\\
020903 &0.251  &8.07 &0.0024 $\pm$ 0.0006 &quasi-spherical &0.0024 $\pm$ 0.0006  &18,2,7,19\\
050826 &0.296  &8.83 &0.03$^{+0.04}_{-0.02}$ &$>$0.2 &$>$0.0006 $\pm$ 0.0004  &6,20,21,22\\ 
020819B &0.410  &9.0 &0.68 $\pm$ 0.17 &\nodata &$<$0.68 $\pm$ 0.17 &23,24,7\\
990712 &0.434  &$\sim$8.4$^{+0.2}_{-0.1}$ &0.67 $\pm$ 0.13 &$>$0.71 $\pm$ 0.03 &$>$0.16 $\pm$ 0.04 &25,2,7,26, 27\\
010921 &0.451  &8.24 &0.95 $\pm$ 0.10 &0.56 $\pm$ 0.04 &0.15 $\pm$ 0.04 &28,20,7,26, 29\\
020405 &0.691  &8.33/8.59 &10 $\pm$ 1 &0.14 $\pm$ 0.02 &0.10 $\pm$ 0.04 &30,2,7,26, 31\\ 
991208 &0.706  &8.02 &22.3 $\pm$ 1.8 &$<$0.15 $\pm$ 0.01 &$<$0.25 $\pm$ 0.06 &32,20,7,26, 33\\
030528 &0.782  &$\sim$8.4$^{+0.2}_{-0.1}$ &2.5 $\pm$ 0.3 &0.14 &0.024 $\pm$ 0.003 &34,20,7,35 \\
051022 &0.807  &8.62 &54 $\pm$ 5 &0.08 &0.17 $\pm$ 0.02 &36,2,7,37\\ 
050824 &0.828  &$\sim$8.4 $\pm$ 0.2 &0.130 $\pm$ 0.029 &0.08 &0.0004 $\pm$ 0.0001 &6,20,3,38\\ 
980703 & 0.966 &8.31/8.65 &7.2 $\pm$ 0.7 &0.20 $\pm$ 0.02 &0.14 $\pm$ 0.01 &39,20,7,26, 40\\ 
\enddata
\tablenotetext{a}{Host metallicity, derived from the Kobulnicky \& Kewley (2004) R$_{23}$ diagnostic. Metallicities have a systematic error of $\pm$0.1 dex except where noted.}
\tablenotetext{b}{References: (1) Tinney et al.\ (1998), (2) Levesque et al.\ (2010a), (3) Amati (2006), (4) Kulkarni et al.\ (1998), (5) Li \& Chevalier (1999), (6) Swift data archive, (7) Amati et al.\ (2008), (8) Soderberg et al.\ (2006), (9) Sakamoto et al.\ (2010), (10) Chornock et al.\ (2010), (11) Starling et al.\ (2010), (12) Soderberg et al. (2010b), (13) Mereghetti \& Gotz (2003), (14) Soderberg et al. (2004b), (15) Greiner et al.\ (2003), (16) Berger et al. (2003), (17) Frail et al. (2005), (18) Ricker et al.\ (2002), (19) Soderberg et al.\ (2004a), (20) Levesque et al.\ (2010b), (21) Butler et al.\ (2007), (22) Mirabal et al. (2007), (23) Hurley et al.\ (2002), (24) Levesque et al.\ (2010c), (25) Galama et al.\ (1999), (26) Bloom et al.\ (2003), (27) Fruchter et al.\ (2000), (28) Djorgovski et al.\ (2001), (29) Price et al.\ (2002), (30) Masetti et al.\ (2002),  (31) Price et al.\ (2003), (32) Hurley \& Cline (1999), (33) Sagar et al. (2000), (34) Atteia et al.\ (2003), (35) Rau et al.\ (2004) (36) Hurley et al.\ (2005), (37) Nakagawa et al.\ (2006), (38) Racusin et al.\ (2009), (39) Djorgovski et al.\ (1998), (40) Frail et al.\ (2003)}
\end{deluxetable}

\begin{references}
\reference {} Amati, L. 2006, MNRAS, 372, 233
\reference {} Amati, L., Guidorzi, C., Frontera, F., Della Valle, M., Finelli, F., Landi, R., \& Montanari, E. 2008, MNRAS, 391, 577
\reference {} Atteia et al.\ 2003, GRB Coordinates Network, 2256, 1
\reference {} Berger, E., et al.\ 2003, Nature, 426, 154
\reference {} Berger, E., Fox, D. B., Kulkarni, S. R., Frail, D. A., \& Djorgovski, S. G. 2007, ApJ, 660, 504
\reference {} Bloom, J. S., Frail, D. A., \& Kulkarni, S. R. 2003, ApJ, 594, 674
\reference {} Bucciantini, N., Quataert, E., Metzger, B. D., Thompson, T. A., Arons, J., \& Del Zanna, L. 2009, MNRAS, 396, 2038
\reference {} Burrows, A., Dessart, L., Livne, E., Ott, C. D., \& Murphy, J. 2007, ApJ, 664, 416
\reference {} Butler, N. R., Kocevski, D., Bloom, J. S., \& Curtis, J. L. 2007, ApJ, 671, 656
\reference {} Chornock, R. et al.\ 2010, ApJL, submitted (arXiv:1004.2262)
\reference {} Crowther, P. A., Dessart, L., Hillier, D. J., Abbott, D. B., \& Fullerton, A. W. 2002, A\&A, 392, 653 
\reference {} Dessart, L., Burrows, A., Livne, E., \& Ott, C. D. 2008, ApJ, 673, 43
\reference {} Djorgovski, S. G., Kulkarni, S. R., Goodrich, R., Frail, D. A., \& Bloom, J. S. 1998, GRB Coordinates Network, 137, 1
\reference {} Djorgovski, S. G., et al.\ 2001, GRB Coordinates Network, 1108, 1
\reference {} Frail, D. A., et al.\ 2001, ApJ, 562, 55
\reference {} Frail, D. A., et al.\ 2003, ApJ, 590, 992
\reference {} Frail, D. A., Soderberg, A. M., Kulkarni, S. R., Berger, E., Yost, S., Fox, D. W., \& Harrison, F. A. 2005, ApJ, 619, 994
\reference {} Fruchter, A. S. et al.\ 2006, Nature, 441, 463
\reference {} Fruchter, A., Vreeswijk, P., Hook, R., \& Pian, E. 2000, GRB Coordinates Network, 752, 1
\reference {} Fryer, C. L. \& Heger, A. 2005, ApJ, 623, 302
\reference {} Galama, T. J., et al.\ 1999, GRB Coordinates Network, 388, 1
\reference {} Ghirlanda, G., Ghisellini, G., \& Lazzati, D. 2004, ApJ, 616, 331
\reference {} Gorosabel, J., et al.\ 2005, A\&A, 444, 711
\reference {} Graham, J. F., Fruchter, A. S., Kewley, L. J., Levesque, E. M., Levan, A. J., Tanvir, N. R., Reichart, D. E., \& Nysewander, M. 2009, in AIP Conf. Proc. Ser. 1133, Gamma-ray Burst: 6th Huntsville Symp., ed. C. Meegan, C. Kouveliotou, \& N. Gehrels (Melville, NY: AIP), 269
\reference {} Greiner, J., Peimbert, M., Estaban, C., Kaufer, A., Jaunsen, A., Smoke, J., Klose, S., \& Reimer, O. 2003, GRB Coordinates Network, 2020, 1
\reference {} Hammer, F., Flores, H., Schaerer, D., Dessauges-Zavadsky, M., Le Floc'h, E., \& Puech, M. 2006, A\&A, 454, 103
\reference {} Henry, R. B. C. \& Worthey, G. 1999, PASP, 111, 919
\reference {} Hirschi, R., Meynet, G., \& Maeder, A. 2005, A\&A, 443, 581
\reference {} Hunter, I., et al.\ 2007, A\&A, 466, 277
\reference {} Hurley, K. \& Cline, T. 1999, GRB Coordinates Network, 450, 1
\reference {} Hurley et al.\ 2002, GRB Coordinates Network, 1507, 1
\reference {} Hurley et al.\ 2005, GRB Coordinates Network, 4139, 1
\reference {} Kewley, L. J., Brown, W. R., Geller, M. J., Kenyon, S. J., \& Kurtz, M. J. 2007, AJ, 133, 882
\reference {} Kewley, L. J. \& Dopita, M. A. 2002, ApJS, 142, 35
\reference {} Kewley, L. J. \& Ellison, S. L. 2008, ApJ, 681, 1183
\reference {} Kobulnicky, H. A. \& Kewley, L. J. 2004, ApJ, 617, 24
\reference {} Kocevski, D., West, A. A., \& Modjaz, M. 2009, ApJ, 702, 377
\reference {} Kulkarni, S. R., et al.\ 1998, Nature, 395, 663
\reference {} Levesque, E. M., Berger, E., Kewley, L. J., \& Bagley, M. M. 2010a, AJ, 139, 694
\reference {} Levesque, E. M., Kewley, L. J., Berger, E., \& Zahid, H. J. 2010b, AJ, in press (arXiv:1006.3560)
\reference {} Levesque, E. M., Kewley, L. J., Graham. J. F., \& Fruchter, A. S. 2010c, ApJ, 712, L26
\reference {} Levesque, E. M., et al.\ 2010d, ApJ, 709, L26
\reference {} Li, Z.-Y. \& Chevalier, R. A. 1999, ApJ, 526, 716
\reference {} MacFadyen, A. I. \& Woosley, S. E. 1999, ApJ, 524, 262
\reference {} Masetti, N., Palazzi, E., Pian, E., Hjorth, J., Castro-Tirado, A., Boehnhardt, H., \& Price, P. 2002, GRB Coordinates Network, 1330, 1
\reference {} Mereghetti, S. \& Gotz, D. 2003, GRB Coordinates Network, 2460, 1
\reference {} Meynet, G. \& Maeder, A. 2000, A\&A, 361, 101
\reference {} Meynet, G. \& Maeder, A. 2007, A\&A, 464, L11
\reference {} Mirabal, N., Halpern, J. P., \& O'Brien, P. T. 2007, ApJ, 61, 127
\reference {} Modjaz, M., et al.\ 2006, ApJ, 645, L21
\reference {} Modjaz, M., et al.\ 2008, AJ, 135, 1136
\reference {} Nagataki, S. 2009, ApJ, 704, 937
\reference {} Nakagawa, Y. E., et al.\ 2006, PASJ, 58, 35
\reference {} Podsiadlowski, P., Ivanova, N., Justham, S., \& Rappaport, S. 2010, MNRAS, in press (arXiv:1004.0249)
\reference {} Podsiadlowski, P., Mazzali, P. A., Nomoto, K., Lazzati, D., \& Cappellaro, E. 2004, ApJ, 607, L17
\reference {} Price, P. A., et al.\ 2002, ApJ, 572, L51
\reference {} Price, P. A., et al.\ 2003, ApJ, 589, 838
\reference {} Prochaska, J. X., Chen, H.-W., Dessauges-Zavadsky, M., \& Bloom, J. S. 2007, ApJ, 666, 267
\reference {} Racusin, J. L., et al.\ 2009, ApJ, 698, 43
\reference {} Ramirez-Ruiz, E., Lazzati, D., \& Blain, A. W. 2002, ApJ, 565, L9
\reference {} Rau, A., et al.\ 2004, A\&A, 427, 815
\reference {} Rau, A., et al.\ 2010, ApJ, 720, 862
\reference {} Ricker et al.\ 2002, GRB Coordinates Network, 1530, 1
\reference {} Sakamoto, T., et al.\ 2010, GRB Coordinates Network, 10511, 1
\reference {} Sagar, R., Mohan, V., Pandey, A. K., Pandey, S. B., \& Castro-Tirado, A. J. 2000, BASI, 28, 15
\reference {} Sari, R., Piran, T., \& Halpern, J. P. 1999, ApJ, 519, L17
\reference {} Savaglio, S., Glazebrook, K., \& Le Borgne, D. 2009, ApJ, 691, 182
\reference {} S\`{i}mon-D\`{i}az, S., Herrero, A., Esteban, C., \& Najarro, F. 2006, A\&A, 448, 351
\reference {} Soderberg, A. M., et al.\ 2004a , ApJ, 606, 994
\reference {} Soderberg, A. M., et al.\ 2004b, Nature, 430, 648
\reference {} Soderberg, A. M. 2006, AIPC, 836, 380
\reference {} Soderberg, A. M., et al.\ 2010, Nature, 463, 513
\reference {} Soderberg, A. M. et al.\ 2010b, ApJ, in prep
\reference {} Sollerman, J., et al.\ 2005, New Astronomy, 11, 103
\reference {} Stanek, K. Z., et al.\ 2006, Acta Astron. 56, 333
\reference {} Starling, R. L. C., et al.\ 2010, MNRAS, submitted (arXiv:1004.2919)
\reference {} Tinney, C., Stathakis, R., Cannon, R., \& Galama, T. 1998, IAU Circ., 6896, 1
\reference {} Uzdensky, D. A. \& MacFadyen, A. I. 2007, ApJ, 669, 546
\reference {} van Zee, L., Salzer, J. J., Haynes, M. P., O'Donoghue, A. A., \& Balonek, T. J. 1998, AJ, 116, 2805
\reference {} Wainwright, C., Berger, E., \& Penprase, B. E. 2007, ApJ, 657, 367
\reference {} Wheeler, J. C., Yi, I., Hoflich, P., \& Wang, L. 2000, ApJ, 537, 810
\reference {} Wolf, C. \& Podsiadlowski, P. 2007, MNRAS, 375, 1049
\reference {} Woosley, S. E. 1993, ApJ, 405, 273
\reference {} Woosley, S. E., Heger, A., \& Weaver, T. A. 2002, Rev. Rod. Phys., 74, 1015
\reference {} Woosley, S. E. \& Heger, A. 2006, ApJ, 637, 914
\reference {} Yoon, S.-C., Langer, N., \& Norman, C. 2006, A\&A, 460, 1999
\reference {} Zaritsky, D. , Kennicutt, R. C., \& Huchra, J. P. 1994, ApJ, 420, 87
\end{references}
\end{document}